\documentclass{article}

\begin{document}

\newcommand{\half}{\frac{1}{2}}

\thispagestyle{empty}

\begin{center}
 {\Large
    Comments on Noncommutative ADHM Construction
 }
\end{center}

\vspace*{2cm}
\begin{center}
 \noindent
 {\large Yu Tian}
 \vspace{5mm}
 \noindent
 \hspace{0.7cm} \parbox{120mm}{\it
School of Physics, Peking University, Beijing 100871, China
\\
E-mail: {\tt phytian@yahoo.com}
 }
\end{center}
 \vspace{5mm}
\begin{center}
 \noindent
 {\large Chuan-Jie Zhu}
 \vspace{5mm}
 \noindent
 \hspace{0.7cm} \parbox{120mm}{\it
Institute of Mathematics, Henan University, Kaifeng 475001
\\
and
\\
Institute for Theoretical Physics, Chinese Academy of Sciences
\\
P. O. Box 2735, Beijing 100080
\\
E-mail: {\tt zhucj@itp.ac.cn}
 }
\end{center}

\vspace{2cm}
\begin{center}
\bf{Abstract}
\end{center}
We extend the method of matrix partition to obtain explicitly the
gauge field for noncommutative ADHM construction in some general
cases. As an application of this method we apply it to the $U(2)$
2-instanton and get explicit result for the gauge fields in the
coincident instanton limit. We also easily apply it to the
noncommutative 't Hooft instantons in the appendix.

\newpage

\section{Introduction}

In a previous paper \cite{TianZhu}, we studied the $U(1)$ and
$U(2)$ instanton solutions on general noncommutative $\bf{R}^4$
\cite{Schwarz}, where for the $U(1)$ 1-instanton, 2-instanton and
$U(2)$ 1-instanton cases we obtain explicitly the gauge field by
ADHM construction \cite{ADHM}.

In (noncommutative) ADHM construction, there are two steps to
obtain the instanton configuration (the gauge field) explicitly.
The first step is to solve the ADHM equation to get the matrix
$\Delta$ (we follow closely the notations of \cite{TianZhu}). The
second step is to construct the matrix $U$ from $\Delta$, or in
other words, to find all the zero modes of $\Delta$. The gauge
field can then be constructed from $U$. The difficulty of the
first step has been well-known for a long time. But it was only
recently realized that the second step is also a difficult problem
in the noncommutative case. \cite{Paperc} discussed the $U(1)$
$\theta_2>0$ case, and our previous paper \cite{TianZhu} resolved
the problem in some of the $U(1)$ and $U(2)$ cases (see also
\cite{Papera,Paperb,Paperd}), but for the general cases there were
no systematic methods to obtain the matrix $U$ yet. It would be a
formidable task to find the matrix $U$ for higher $N$ and $k$.

Keeping this problem in mind, we will try to develop a method to
accomplish the second step  systematically. In fact there is a
general method in the commutative case \cite{decomp}. What we will
do is to extend this method to the noncommutative case. By using
this method we have rechecked all the results obtained in our
previous paper \cite{TianZhu}. As a new application of this
method, we will demonstrate how to deal with some of the $U(2)$
2-instanton cases. As far as we know there is no general solution
of noncommutative $U(2)$ 2-instanton ADHM equation. So we will use
this method to study a special case in the coincident instanton
limit.

This paper is organized as follows: in section 2 we recall briefly
the noncommutative $\bf{R}^4$ and set our notations and then the
ADHM construction. In section 3 we present the modified matrix
partition method and in section 4 we solve a special case of
$U(2)$ 2-instanton. Finally we mention that our method can be used
to give a more general class of $U(2)$ multi-instantons.

\section{$\bf{R}_{\rm  NC}^4$,  the (anti-)self-dual equations and
ADHM construction}

\subsection{$\bf{R}_{\rm  NC}^4$ and  the (anti-)self-dual
equations}

First let us recall briefly the noncommutative $\bf{R}^4$ and set
our notations\footnote{For general reviews on noncommutative
geometry and field theory, see, for example, \cite{Paperc,
Reviewa, Reviewb, Reviewc}.}. For a general noncommutative
$\bf{R}^4$ we mean a space with (operator) coordinates $x^m$,
$m=1, \cdots, 4$, which satisfy the following relations:
\begin{equation}
[x^m,x^n] = i \theta^{mn},
\end{equation}
where $\theta^{mn}$ are real constants. If we assume the standard
(Euclidean) metric for the noncommutative $\bf{R}^4$, we can use
the orthogonal transformation with positive determinant to change
$\theta^{mn}$ into the following standard form:
\begin{equation}
\label{theta} (\theta^{mn})=\left(\begin{array}{cccc} 0 &
\theta^{12} & 0 & 0 \\ -\theta^{12}& 0 & 0 & 0 \\ 0 & 0 & 0
&\theta^{34}\\ 0 & 0 & -\theta^{34}& 0 \end{array}\right),
\end{equation}
with $\theta^{12}>0$ and $\theta^{12}+\theta^{34}\geq 0$. If the
original $\theta^{mn}$ is non-degenerate, $\theta^{34}$ in this
standard form is non-vanishing. Otherwise the considered
noncommutative $\bf{R}^4$ can be identified with a direct product
$\bf{R}_{\rm  NC}^2\times\bf{R}^2$. One can refer \cite{degen} for
this degenerate case, which we will not dwell on in this paper. By
using this form of $\theta^{mn}$, the only non-vanishing
commutators are as follows:
\begin{equation}
[x^1,x^2] = i \theta^{12}, \qquad [x^3,x^4] = i \theta^{34},
\end{equation}
and the other twos obtained by using the anti-symmetric property
of commutators. Introducing complex coordinates:
\begin{equation}
\label{complex}
\begin{array}{rl}
z_1 = x^2 + i x^1 , & \bar{z}_1 = x^2 - i x^1, \\
z_2 = x^4 + i x^3 , & \bar{z}_2 = x^4 - i x^3,
\end{array}
\end{equation}
the non-vanishing commutation relations are
\begin{equation}
\label{commutator of z} [\bar
z_1,z_1]=2\theta^{12}\equiv\theta_1,\quad [\bar z_2,z_2]
=2\theta^{34}\equiv\theta_2.
\end{equation}

By a noncommutative gauge field $A_m$ we mean an operator valued
field. The (anti-hermitian) field strength $F_{mn}$ is defined
similarly as in the commutative case:
\begin{equation}
\label{F by A} F_{mn}=\hat\partial_{[m} A_{n]} + A_{[m} A_{n]}
\equiv \hat\partial_m A_n - \hat\partial_n A_m + [A_m, A_n],
\end{equation}
where the derivative operator $\hat\partial_m$ is defined as
follows:
\begin{equation}
\hat\partial_m f \equiv - i \theta_{mn} [x^n, f],
\end{equation}
where $\theta_{mn}$ is the inverse of $\theta^{mn}$. For our
standard form (\ref{theta}) of $\theta^{mn}$ we have
\begin{equation}
\hat\partial_1 f = {i \over \theta^{12}} [x^2,f], \qquad
\hat\partial_2 f = - {i \over \theta^{12}} [x^1,f],
\end{equation}
which can be expressed by the complex coordinates (\ref{complex})
as follows:
\begin{equation}
\partial_1 f \equiv \hat\partial_{z_1}f = {1\over\theta_1}
[\bar z_1, f], \qquad \bar\partial_1 f \equiv \hat\partial_{\bar
z_1}f = - {1\over\theta_1}[z_1, f],
\end{equation}
and similar relations for $x^{3,4}$ and $z_2,\bar{z}_2$.

For a general metric $g_{mn}$ the instanton equations are
\begin{equation}
\label{instanton}
F_{mn}=\pm\frac{\epsilon^{pqrs}}{2\sqrt{g}}g_{mp}g_{nq}F_{rs},
\end{equation}
and the solutions are known as self-dual (SD, for ``+'' sign) and
anti-self-dual (ASD, for ``$-$'' sign) instantons. Here
$\epsilon^{pqrs}$ is the totally anti-symmetric tensor
($\epsilon^{1234}=1$ etc.) and $g$ is the metric. We will take the
standard metric $g_{mn}=\delta_{mn}$ and take the noncommutative
parameters $\theta_{1,2}$ as free parameters. We also note that
the notions of self-dual and anti-self-dual are interchanged by a
parity transformation. A parity transformation also changes the
sign of $\theta^{mn}$. In the following discussion we will
consider only the ASD instantons. So we should not restrict
$\theta_2$ to be positive.

\subsection{ADHM construction for ordinary gauge theory}

For ordinary gauge theory all the (ASD) instanton solutions are
obtained by ADHM (Atiyah-Drinfeld-Hitchin-Manin) construction
\cite{ADHM}. In this construction we introduce the following
ingredients (for $U(N)$ gauge theory with instanton number $k$):
\begin{itemize}
\item complex vector spaces $V$ and $W$ of dimensions $k$ and $N$,
\item $k\times k$ matrix $B_{1,2}$, $k\times N$ matrix $I$ and
$N\times k$ matrix $J$,
\item the following quantities:
\begin{eqnarray}
\label{ADHM1} \mu_r & = & [B_1, B_1^\dagger] + [B_2, B_2^\dagger]
+ I \,
I^\dagger -J^\dagger J, \\
\label{ADHM2} \mu_c & = & [B_1,B_2] + I\, J.
\end{eqnarray}
\end{itemize}
The claim of ADHM is as follows:
\begin{itemize}
\item Given $B_{1,2}$, $I$ and $J$ such that $\mu_r=\mu_c=0$, an
ASD gauge field can be constructed;
\item All ASD gauge fields can be obtained in this way.
\end{itemize}

It is convenient to introduce a quaternionic notation for the
4-dimensional Euclidean space-time indices:
\begin{equation}
x\equiv x^n\sigma_n,\qquad\bar{x}\equiv x^n\bar\sigma_n,
\end{equation}
where $\sigma_n=(i\vec{\tau},1)$ and $\tau^c$, $c=1,2,3$ are the
three Pauli matrices, and the conjugate matrices
$\bar\sigma_n=\sigma_n^\dag=(-i\vec{\tau},1)$. In terms of the
complex coordinates (\ref{complex}) we have
\begin{equation}
(x_{\alpha\dot\alpha})=\left(\begin{array}{cc} z_2 & z_1 \\ -
\bar{z}_1 & \bar{z}_2 \end{array}\right), \qquad
(\bar{x}^{\dot\alpha\alpha})=\left(\begin{array}{cc} \bar{z}_2 & -
z_1 \\ \bar{z}_1 & z_2 \end{array}\right).
\end{equation}
Then the basic object in the ADHM construction is the
$(N+2k)\times 2k$ matrix $\Delta$ which is linear in the
space-time coordinates:
\begin{equation}
\label{Delta} \Delta=a+b\bar{x},
\end{equation}
where the constant matrices
\begin{equation}
a = \left( \begin{array}{cc} I^\dag & J \\ B_2^\dagger & -B_1
\\ B_1^\dagger & B_2 \end{array} \right), \quad
b = \left( \begin{array}{cc} 0 & 0 \\ 1 & 0 \\ 0 & 1 \end{array}
 \right).
\end{equation}

Consider the conjugate operator of $\Delta$:
\begin{equation}
\Delta^\dagger = a^\dag +x b^\dag = \left( \matrix{I & B_2 + z_2 &
B_1 + z_1 \cr J^\dagger & -B_1^\dagger -\bar z_1 & B_2^\dagger +
\bar z_2} \right).
\end{equation}
It is easy to check that the ADHM equations (\ref{ADHM1}) and
(\ref{ADHM2}) are equivalent to the so-called factorization
condition:
\begin{equation}
\label{factorize} \Delta^\dagger\Delta=\left(\matrix{f^{-1} & 0
\cr 0 & f^{-1}}\right),
\end{equation}
where $f(x)$ is a $k\times k$ hermitian matrix. From the above
condition we can construct a hermitian projection operator $P$ as
follows:\footnote{We use the following abbreviation for
expressions with $f$:
\begin{equation}
\Delta f \Delta^\dag \equiv \Delta \left(\matrix{f & 0 \cr 0 &
f}\right) \Delta^\dag =\Delta (f\otimes 1_2)\Delta^\dag.
\end{equation}
}
\begin{eqnarray}
P&=&\Delta f\Delta^\dag, \cr P^2&=&\Delta f f^{-1}f\Delta^\dag=P.
\end{eqnarray}

Obviously, the null-space of $\Delta^\dagger(x)$ is of $N$
dimension for generic $x$. The basis vectors for this null-space
can be assembled into an $(N+2k)\times N$ matrix $U(x)$:
\begin{equation}
\Delta^\dag U=0,
\end{equation}
which can be chosen to satisfy the following orthonormal
condition:
\begin{equation}
\label{normal} U^\dag U=1.
\end{equation}
The above orthonormal condition guarantees that $UU^\dag$ is also
a hermitian projection operator. Now it can be proved that the
completeness relation \cite{Paperb}
\begin{equation}
\label{complete} P+UU^\dag=1
\end{equation}
holds if $U$ contains the whole null-space of $\Delta^\dagger$. In
other words, this completeness relation requires that $U$ consists
of all the zero modes of $\Delta^\dagger$.\footnote{The proof is
sketched as follows: the two projection operators $P$ and
$UU^\dag$ are orthogonal to each other, and so $1-P-UU^\dag$ is
also a hermitian projection operator. Now this can always be
written as the form $VV^\dag$. Here $\Delta$ and $f$ are both of
maximum rank and $PVV^\dag=0$, then $V$ must consist of some zero
modes of $\Delta^\dagger$ other than those in $U$. This conclusion
is in conflict with the assumption that $U$ contains all the zero
modes of $\Delta^\dagger$. Although the notion of maximum rank is
ambiguous in the infinite-dimensional case as we will encounter in
noncommutative ADHM construction, it can be defined as follows so
that the above proof is also true in the infinite-dimensional
case. We say that an $\infty\times\infty$ matrix $X$ is of maximum
rank if it has no zero modes, i.e., $X\Phi\ne 0$ if $\Phi\ne 0$.}

The (anti-hermitian) gauge potential is constructed from $U$ by
the following formula:
\begin{equation}
A_m= U^\dag\partial_m U.
\end{equation}
Substituting this expression into (\ref{F by A}), we get the
following field strength:
\begin{eqnarray}
\label{calculate F} F_{mn}&=&\partial_{[m}(U^\dag\partial_{n]}U)
+(U^\dag\partial_{[m}U)(U^\dag\partial_{n]}U)
=\partial_{[m}U^\dag(1-UU^\dag)\partial_{n]}U\nonumber\\
&=&\partial_{[m}U^\dag\Delta f\Delta^\dag\partial_{n]}U
=U^\dag\partial_{[m}\Delta f\partial_{n]}\Delta^\dag U
=U^\dag b\bar\sigma_{[m}\sigma_{n]}f b^\dag U\nonumber\\
&=& 2i\bar\eta^c_{mn}U^\dag b(\tau^c f)b^\dag U.
\end{eqnarray}
Here $\bar\eta^a_{ij}$ is the standard 't Hooft $\eta$-symbol,
which is anti-self-dual:
\begin{equation}
\half\epsilon_{ijkl}\bar\eta^a_{kl}=-\bar\eta^a_{ij}.
\end{equation}

\subsection{Noncommutative ADHM construction}

The above construction has been extended to noncommutative gauge
theory \cite{Schwarz}. We recall this construction briefly here.
By introducing the same data as above but considering the $z_i$'s
as noncommutative we see that the factorization condition
(\ref{factorize}) still gives $\mu_c=0$, but $\mu_r$ no longer
vanishes. It is easy to check that the following relation holds:
\begin{equation} \mu_r=\zeta\equiv\theta_1+\theta_2.
\end{equation}
In this case the two ADHM equations (\ref{ADHM1}) and
(\ref{ADHM2}) can be combined into one \cite{Paperd}:
\begin{equation}
\label{ADHM} \tau^{c\dot\alpha}{}_{\dot\beta}(\bar a^{\dot\beta}
a_{\dot\alpha})_{ij}=\delta_{ij}\delta^{c3}\zeta.
\end{equation}

As studied mathematically by various people (see, for example, the
lectures by H. Nakajima \cite{Nakajima}), the moduli space of the
noncommutative instantons is better behaved than their commutative
counterpart. In the noncommutative case the operator
$\Delta^\dagger\Delta$ always has maximum rank (see
\cite{Reviewa}).

Though there is no much difference between the noncommutative ADHM
construction and the commutative one, we should study the
noncommutative case in more detail. In order to study the
instanton solutions precisely, we use a Fock space representation
as follows ($n_1, n_2\geq 0$):
\begin{eqnarray}
z_1|n_1,n_2\rangle & = &
\sqrt{\theta_1}\sqrt{n_1+1}|n_1+1,n_2\rangle,
\\
\quad\bar z_1|n_1,n_2\rangle & =&
\sqrt{\theta_1}\sqrt{n_1}|n_1-1,n_2 \rangle,
\end{eqnarray}
by using the commutation relation (\ref{commutator of z}). Similar
expressions for $z_2$ and $\bar z_2$ also apply (but paying a
little attention to the sign of $\theta_2$ which is not restricted
to be positive). In this representation the $z_i$'s are
infinite-dimensional matrices, and so are the operator $\Delta$,
$\Delta^\dag$ etc. Because of infinite dimensions are involved we
can not determine the dimension of null-space of $\Delta^\dag$
straightforwardly from the difference of the numbers of its rows
and columns. But it turns out that $\Delta^\dag$ also has infinite
number of zero modes, and they can be arranged into an
$(N+2k)\times N$ matrix with entries from the (noncommutative)
algebra generated by the coordinates, which resembles the
commutative case.

\section{The modified matrix partition method}

In the commutative case, there is a standard method \cite{decomp},
which we call the `matrix partition method' in ADHM construction,
to obtain the matrix $U$ from $\Delta$ as follows. First we
introduce the following decomposition of the matrices $U$ and
$\Delta$:
\begin{equation}\label{block}
U=\left(\begin{array}{c} V_{N\times N} \\
U'_{2k\times N}
\end{array}\right),\qquad
\Delta=\left(\begin{array}{c} K_{N\times 2k} \\ \Delta'_{2k\times
2k} \end{array}\right).
\end{equation}
So we have
\begin{equation}\label{Deltap}
K=\left(\begin{array}{cc} I^\dag & J
\end{array}\right),\quad \Delta'=\left(\begin{array}{cc}
B_2^\dag+\bar{z}_2 & -B_1-z_1
\\ B_1^\dag+\bar{z}_1 & B_2+z_2 \end{array}\right).
\end{equation}
Substituting (\ref{block}) into the completeness relation
(\ref{complete}), we obtain
\begin{equation}
\label{block complete}
\left(\begin{array}{cc} V V^\dag & V U'^\dag \\
U'V^\dag & U'U'^\dag \end{array}\right)=\left(\begin{array}{cc}
1-K f K^\dag & -K f \Delta'^\dag \\ -\Delta'f K^\dag &
1-\Delta'f\Delta'^\dag
\end{array}\right).
\end{equation}
Now we can choose
\begin{equation}
\label{singular} V=V^\dag=(1-K f K^\dag)^{1/2}
\end{equation}
as a solution of the (1,1) element of the matrix equation
(\ref{block complete}). Then by the (2,1) element of (\ref{block
complete}) the matrix $U'$ can be expressed as
\begin{equation}
\label{Up} U'=-\Delta'f K^\dag V^{-1}.
\end{equation}
The choice (\ref{singular}) of $V$ is known as the `singular
gauge', and any other choices of $V$ which solve (\ref{block
complete}) are related to (\ref{singular}) by a gauge
transformation.

Corresponding to the singularity arising in the commutative case,
the choice (\ref{singular}) also brings problems if we extend the
matrix partition method to the noncommutative case. In the
following subsections we will show that slight modifications of
this method can remedy these problems.

In the noncommutative case, the solutions of the (1,1) element of
(\ref{block complete}) are not always related to each other by
gauge transformations. This can be seen as follows.  If $V$ is a
solution of the (1,1) element of (\ref{block complete}) and $S$ is
an operator such that $SS^\dag=1$, then $VS$ is also a solution:
\begin{equation}
(VS)(VS)^\dag=VSS^\dag V^\dag=VV^\dag,
\end{equation}
while we can have $S^\dag S\ne 1$. This $S$ induces a new solution
$US$ of the completeness relation (\ref{complete}). However, $US$
does not satisfy the orthonormal condition (\ref{normal}):
\begin{equation}
(US)^\dag(US)=S^\dag U^\dag US\ne 1,
\end{equation}
which is not a problem in the commutative case: any solution of
(\ref{complete}) must satisfy (\ref{normal}). The above freedom of
$V$ can be used to resolve the problems of the matrix partition
method in noncommutative ADHM construction, while we must consider
the completeness relation (\ref{complete}) and the orthonormal
condition (\ref{normal}) as two independent equations.

The two cases, $\theta_2>0$ and $\theta_2<0$, have an essential
difference: for $\theta_2>0$ the matrix $\Delta'$ has zero modes
and $\Delta'^\dag$ is of maximum rank, while for $\theta_2<0$ the
matrix $\Delta'$ is of maximum rank and $\Delta'^\dag$ has zero
modes. As we can see in the following subsections, the two cases
must be considered separately just because of this difference.

\subsection{$\theta_2>0$ case}
The immediate consequence of that $\Delta'$ has zero modes in this
case is that the matrix $V$ in (\ref{singular}) is not invertible
and we we can not use (\ref{Up}) to determine $U'$. Here we will
try to directly project out the zero modes of $VV^\dag$ first.

Let $\Pi_0$ be the projector whose range is the null-space of
$(1-KfK^\dag)$. We can still use (\ref{singular}) and let
\begin{equation}
\label{Upp} U'=-\Delta'f K^\dag(1-K f K^\dag)^{-1/2}_\Pi,
\end{equation}
where the inversion of $(1-KfK^\dag)$ is restricted on $\Pi\equiv
1-\Pi_0$. Noting that
\begin{equation}
\begin{array}{rl}
\Delta'^\dag\Delta'fK^\dag\Pi_0&=\Delta'^\dag\Delta'(K^\dag
K+\Delta'^\dag\Delta')^{-1}K^\dag\Pi_0\\
&=[1-K^\dag K(K^\dag K+\Delta'^\dag\Delta')^{-1}]K^\dag\Pi_0\\
&=K^\dag(1-K f K^\dag)\Pi_0=0
\end{array}
\end{equation}
and  $\Delta'^\dag$ is of maximum rank by (\ref{Deltap}), we have
\begin{equation}\label{zero}
\Delta'fK^\dag\Pi_0=0.
\end{equation}
The (2,1) element of the matrix equation (\ref{block complete}) is
now satisfied:
\begin{equation}
\begin{array}{rl}
U'V^\dag&=-\Delta'f K^\dag(1-K f K^\dag)^{-1/2}_\Pi(1-K f
K^\dag)^{1/2}\\
&=-\Delta'f K^\dag(1-\Pi_0)=-\Delta'f K^\dag.
\end{array}
\end{equation}
Moreover we can easily check that the (2,2) element of (\ref{block
complete}) holds:
\begin{equation}\label{bc22l}\begin{array}{rl}
\Delta'^\dag U'U'^\dag&=\Delta'^\dag\Delta'fK^\dag(1-K
fK^\dag)^{-1}_\Pi Kf\Delta'^\dag\\
&=K^\dag(1-K f K^\dag)(1-K fK^\dag)^{-1}_\Pi Kf\Delta'^\dag\\
&=K^\dag(1-\Pi_0)Kf\Delta'^\dag=K^\dag Kf\Delta'^\dag
\end{array}\end{equation}
and
\begin{equation}\label{bc22r}\begin{array}{rl}
\Delta'^\dag(1-\Delta'f\Delta'^\dag)&=[1-\Delta'^\dag\Delta'
(K^\dag K+\Delta'^\dag\Delta')^{-1}]\Delta'^\dag\\
&=K^\dag Kf\Delta'^\dag,
\end{array}\end{equation}
so
\begin{equation}\label{bc22}
U'U'^\dag=1-\Delta'f\Delta'^\dag.
\end{equation}

Even though (\ref{singular}) together with (\ref{Upp}) satisfy
equation (\ref{block complete}), they do not satisfy the
orthonormal condition (\ref{normal}):
\begin{equation}\begin{array}{rl}
U^\dag U&=V^\dag V+U'^\dag U'\\
&=1-K f K^\dag+(1-K fK^\dag)^{-1/2}_\Pi Kf\Delta'^\dag\Delta'f
K^\dag(1-K f K^\dag)^{-1/2}_\Pi\\
&=1-K f K^\dag\\
&\quad+(1-K fK^\dag)^{-1/2}_\Pi Kf(f^{-1}-K^\dag K)f
K^\dag(1-K f K^\dag)^{-1/2}_\Pi\\
&=1-K f K^\dag\\
&\quad+(1-K fK^\dag)^{-1/2}_\Pi(KfK^\dag-KfK^\dag KfK^\dag)(1-K f
K^\dag)^{-1/2}_\Pi\\
&=1-K f K^\dag+\Pi K f K^\dag\Pi\\
&=1-\Pi_0 K f K^\dag-K f K^\dag\Pi_0+\Pi_0=\Pi\ne 1.
\end{array}\end{equation}
The solution to this problem is to introduce a shift operator $s$
as follows:
\begin{equation}
ss^\dag=1,\quad s^\dag s=\Pi,\label{shift}
\end{equation}
and we set
\begin{eqnarray}
\label{V}V&=&(1-K f K^\dag)^{1/2}s^\dag,\\
\label{U p}U'&=&-\Delta'fK^\dag(1-KfK^\dag)^{-1/2}s^\dag.
\end{eqnarray}
The operator $s^\dag$ removes the zero modes of the operator
$(1-KfK^\dag)$ and makes its inversion well-defined. Then we can
easily see that the completeness relation (\ref{block complete})
still holds and the orthonormal condition is now satisfied:
\begin{equation}\begin{array}{rl}
U^\dag U&=s(1-K f K^\dag)s^\dag\\
&\quad+s(1-K fK^\dag)^{-1/2}Kf\Delta'^\dag\Delta'f K^\dag(1-K f
K^\dag)^{-1/2}s^\dag\\
&=s\Pi s^\dag=1.
\end{array}\end{equation}
So (\ref{V}) and (\ref{U p}) constitute the required matrix $U$.

Now we determine the null-space of $(1-K f K^\dag)$. As
\begin{equation}\label{K zero}
\begin{array}{rl}
(1-KfK^\dag)K&=K[1-(K^\dag K+\Delta'^\dag\Delta')^{-1}K^\dag K]\\
&=Kf\Delta'^\dag\Delta',
\end{array}
\end{equation}
we can first determine the null-space of $\Delta'$. It is easy to
see from (\ref{Deltap}) that the zero modes of $\Delta'$ must take
the form:
\begin{equation}
\Phi=\left(\begin{array}{c}
\Upsilon\\
0\end{array}\right)
\end{equation}
because the second column of $\Delta'$ is of maximum rank. By
(\ref{K zero}), we introduce an operator
\begin{equation}\label{Psi}
\Psi\equiv K\Phi=I^\dag\Upsilon,
\end{equation}
which belongs to the null-space of $(1-KfK^\dag)$:
\begin{equation}
(1-KfK^\dag)\Psi=Kf\Delta'^\dag\Delta'\Phi=0.
\end{equation}
Then a hermitian operator $G$ can be constructed as follows:
\begin{equation}
G=\Psi^\dag\Psi=\Upsilon^\dag II^\dag\Upsilon
=\Upsilon^\dag[II^\dag+
\sum_{\alpha=1,2}(B_\alpha+z_\alpha)(B_\alpha^\dag+\bar{z}_\alpha)]\Upsilon
=\Upsilon^\dag f^{-1}\Upsilon,
\end{equation}
which is positive definite because $f^{-1}$ has no zero mode. Here
we have used the fact
\begin{equation}
(B_\alpha^\dag+\bar{z}_\alpha)\Upsilon=0,\quad\alpha=1,2.
\end{equation}
If $\Psi$ contains all the zero modes of $(1-K f K^\dag)$, it is
easy to show that
\begin{equation}
\Pi_0=\Psi G^{-1}\Psi^\dag.
\end{equation}
We will prove this completeness of $\Psi$ as follows. From
(\ref{zero}) we have
\begin{eqnarray}
\label{fI}fI\Pi_0&\subset&\Upsilon,\\
\label{J}J^\dag\Pi_0&=&0,
\end{eqnarray}
where the notation `$\subset$' now means `included in the space
spanned by the column vectors of', and (\ref{J}) gives
\begin{equation}
(1-KfK^\dag)\Pi_0=(1-I^\dag fI)\Pi_0=0,
\end{equation}
which leads to
\begin{equation}
\Pi_0=I^\dag fI\Pi_0\subset I^\dag\Upsilon.
\end{equation}
So (51) does span the whole null-space of $(1-KfK^\dag)$.

For $[B_1,B_2]\ne 0$, we can not give any more explicit results.
But when $[B_1,B_2]=0$, we have \cite{Paperc}:
\begin{equation}
\Upsilon= e^{-\sum_\alpha\theta_\alpha^{-1}B_\alpha^\dag
z_\alpha}|0,0\rangle.
\end{equation}
In general, the condition $[B_1,B_2]=0$ can not be satisfied. When
$N=1$ it can be proved \cite{Nakajima} that we can always take
$J=0$, so the above condition holds. Moreover, it also holds in
the $U(2)$ 1-instanton case, which we had considered in
\cite{TianZhu}. A special $U(2)$ 2-instanton case also satisfies
this condition. We will give the explicit solution in the next
section.

\subsection{$\theta_2<0$ case}

In this case the problem is that $\Delta'^\dag$ has zero modes.
Here $\Delta'$ is of maximum rank as we can see from
(\ref{Deltap}), so is $(1-KfK^\dag)$ for the reason stated in the
previous subsection. But if we use (\ref{singular}) and (\ref{Up})
as the solution of (\ref{block complete}), equations (\ref{bc22l})
(where $\Pi_0=0$) and (\ref{bc22r}) do not give (\ref{bc22}).

Suppose that the null-space of $\Delta'^\dag$ is of $k'$ dimension
and $\Gamma$ is the matrix composed of all the orthonormal zero
modes of $\Delta'^\dag$:
\begin{equation}
\Gamma^\dag\Gamma=1_{k'\times k'},\qquad\Delta'^\dag\Gamma=0,
\end{equation}
we have
\begin{equation}\label{extra}
1-\Delta'f\Delta'^\dag-U'U'^\dag\subset\Gamma.
\end{equation}
Noting that a matrix $\tilde{U}$ composed of (\ref{singular}) and
(\ref{Up}) satisfies
\begin{equation}
\tilde{U}^\dag\tilde{U}=1,\qquad\Delta^\dag\tilde{U}=0,
\end{equation}
the only thing we must do is to add some extra zero modes to
$\tilde{U}$ so as to make it satisfy the completeness relation. In
what follows, we will prove that these extra zero modes can be
assembled into a matrix $X$ of the form
\begin{equation}
X=\left(\begin{array}{c} 0 \\ \Gamma \end{array}\right) .
\end{equation}
Firstly, $X$ belongs to the null-space of $\Delta^\dag$:
\begin{equation}
\Delta^\dag X=\Delta'^\dag\Gamma=0.
\end{equation}
Secondly, it is not difficult to see that a non-vanishing (1,1)
element of $X$ will break down the (1,1) element of equation
(\ref{block complete}). Finally, any vectors belong to the space
spanned by $\Gamma$ appearing in the (2,1) element of $X$ will
violate (\ref{extra}). So we conclude that the combination of
$\tilde{U}$ and $X$ gives all the zero modes of $\Delta^\dag$.

To make the above conclusion more explicit, we introduce the
following shift operator:
\begin{equation}\label{U begin}
SS^\dag=1,\quad S^\dag S=1-\sum_{n=0}^{k'-1}|n,0\rangle\langle
n,0|
\end{equation}
and a $k'\times N$ matrix
\begin{equation}
Y=\left(\begin{array}{cccc} \langle 0,0| & 0 & \cdots & 0 \\
\langle 1,0| & 0 & \cdots & 0 \\ \vdots & \vdots & \ddots & \vdots
\\ \langle k'-1,0| & 0 & \cdots & 0 \end{array}\right).
\end{equation}
An $N\times N$ matrix operator $\mathcal{S}$ is defined as
\begin{equation}
\mathcal{S}=\mathrm{diag}(S,1,\cdots,1).
\end{equation}
Then the required matrix $U$ can be expressed as follows:
\begin{eqnarray}
V&=&(1-K f K^\dag)^{1/2}\mathcal{S},\\
U'&=&\Gamma
Y-\Delta'fK^\dag(1-KfK^\dag)^{-1/2}\mathcal{S},\label{U end}
\end{eqnarray}
which can be straightforwardly checked to satisfy both conditions
(\ref{normal}) and (\ref{complete}).

Again for $[B_1,B_2]=0$ we can explicitly give
\begin{equation}\label{Gamma}
\Gamma=\left(\begin{array}{c}
e^{\theta_2^{-1}B_2\bar{z}_2-\theta_1^{-1}B_1^\dag z_1}|0,0\rangle
\\ 0
\end{array}\right)W^{-1/2},
\end{equation}
where the normalization factor $W$ is:
\begin{equation}
W=\langle 0,0|e^{\theta_2^{-1}B_2^\dag
z_2-\theta_1^{-1}B_1\bar{z}_1}e^{\theta_2^{-1}B_2
\bar{z}_2-\theta_1^{-1}B_1^\dag z_1}|0,0\rangle,
\end{equation}
and so we have $k'=k$.

\section{A special $U(2)$ 2-instanton solution}

To explicitly give the moduli space of a $U(N)$ multi-instanton is
a very difficult problem \cite{multi}. In the commutative case,
\cite{2-instanton} solved this problem when the instanton number
$k=2$. For the noncommutative case a similar solution has not
appeared yet. However, it is much easier to find some special
solutions of $U(N)$ multi-instanton. To illustrate our method, we
construct one of noncommutative $U(2)$ 2-instanton in the limit of
coincident instantons.

In a special case, these ADHM data (\ref{Delta}) can be written as
follows:
\begin{equation}
\Delta=\left(\begin{array}{cccc} 0 &\sqrt{2\zeta+\rho^2}& 0 & 0\\
0 & 0 & \rho & 0\\
\bar z_2 & 0 & -z_1 & -\sqrt{\zeta+\rho^2}\\
0 & \bar z_2 & 0 & -z_1\\
\bar z_1 & 0 & z_2 & 0\\
\sqrt{\zeta+\rho^2}&\bar z_1 & 0 & z_2 \end{array}\right),
\end{equation}
\begin{equation}
\Delta^\dag=\left(\begin{array}{cccccc} 0 & 0 & z_2 & 0 & z_1 &
\sqrt{\zeta+\rho^2}\\
\sqrt{2\zeta+\rho^2} & 0 & 0 & z_2 & 0 & z_1\\
0 & \rho & -\bar z_1 & 0 & \bar z_2 & 0\\
0 & 0 & -\sqrt{\zeta+\rho^2}& -\bar z_1 & 0 &\bar z_2
\end{array}\right),
\end{equation}
where $\rho$ is the scale of the instanton size. In other words,
we have
\begin{equation}
I=\left(\begin{array}{cc} 0 & 0 \\ \sqrt{2\zeta+\rho^2} & 0
\end{array}\right),
\qquad J=\left(\begin{array}{cc} 0 & 0 \\ \rho & 0
\end{array}\right),
\end{equation}
\begin{equation}
B_1=\left(\begin{array}{cc} 0 & \sqrt{\zeta+\rho^2} \\ 0 & 0
\end{array}\right),
\qquad B_2=0.
\end{equation}
These ADHM data lead to
\begin{equation}
f=\left(\begin{array}{cc}
(Z_1''+Z_2'+\zeta+\rho^2)(Z_\rho''~')^{-1}
& -\sqrt{\zeta+\rho^2}(Z_\rho''~')^{-1}\bar z_1\\
-\sqrt{\zeta+\rho^2}z_1(Z_\rho''~')^{-1}&
(Z_1'+Z_2'-\theta_1+\rho^2)(Z_\rho'~')^{-1}
\end{array}\right).
\end{equation}
In order to make the above expressions more similar to that of the
$U(1)$ 2-instanton discussed in \cite{TianZhu}, we have carefully
chosen our notations as follows:
\begin{equation}
Z_1\equiv z_1\bar{z}_1,\qquad Z_2\equiv z_2\bar{z}_2,
\end{equation}
\begin{eqnarray}
&Z_1'\equiv Z_1+\theta_1,\quad Z_1''\equiv
Z_1+2\theta_1,\quad\rm{etc.}&\\
&Z_2'\equiv Z_2+\theta_2,\quad Z_2''\equiv
Z_2+2\theta_2,\quad\rm{etc.}&
\end{eqnarray}
\begin{eqnarray}
Z&\equiv&(Z_1+Z_2)^2-\theta_1 Z_1+(\theta_2+\rho^2)Z_2,\label{Z U2}\\
Z'~'&\equiv&(Z_1'+Z_2')^2-\theta_1
Z_1'+(\theta_2+\rho^2)Z_2',\quad\rm{etc.}
\end{eqnarray}
\begin{eqnarray}
Z_\rho&\equiv&Z+\rho^2(Z_1+Z_2+\zeta+\rho^2),\\
Z_\rho'~'&\equiv&Z'~'+\rho^2(Z_1'+Z_2'+\zeta+\rho^2),\quad\rm{etc.}
\end{eqnarray}
It is not difficult to obtain
\begin{equation}
1-KfK^\dag=1-\left(\begin{array}{cc} (2\zeta+\rho^2)f_{22} & 0 \\
0 & \rho^2 f_{11} \end{array}\right)=\left(\begin{array}{cc}
Z/Z_\rho'~' & 0 \\
0 & Z''~'/Z_\rho''~' \end{array}\right),
\end{equation}
\begin{eqnarray}
-(\Delta'fK^\dag)_{*1}&=&\sqrt{2\zeta+\rho^2}
\left(\begin{array}{c} \sqrt{\zeta+\rho^2}\bar{z}_2\bar{z}_1 \\
-\bar{z}_2(Z_1+Z_2+\theta_2+\rho^2) \\
\sqrt{\zeta+\rho^2}\bar{z}_1\bar{z}_1 \\
-\bar{z}_1(Z_1+Z_2-\theta_1) \end{array}\right)/Z_\rho'~',\\
-(\Delta'fK^\dag)_{*2}&=&\rho\left(\begin{array}{c} z_1(Z_1''+Z_2') \\
-\sqrt{\zeta+\rho^2}z_1 z_1 \\
-z_2(Z_1''+Z_2'+\zeta+\rho^2) \\
\sqrt{\zeta+\rho^2}z_2 z_1
\end{array}\right)/Z_\rho''~',
\end{eqnarray}
where the subscripts `*1' and `*2' mean the first column and the
second column of the matrix $\Delta' f K^\dag$ respectively. Now
we solve the two cases $\theta_2>0$ and $\theta_2<0$ in sequence.

\subsection{The $\theta_2>0$ case}

Because $[B_1,B_2]=0$ for this special solution, we have
\begin{equation}
\Psi=I^\dag e^{-\sum_\alpha\theta_\alpha^{-1}B_\alpha^\dag
z_\alpha}|0,0\rangle=\sqrt{2\zeta+\rho^2}\left(\begin{array}{cc}
-\sqrt{\frac{\zeta+\rho^2}{\theta_1}}|1,0\rangle & |0,0\rangle \\
0 & 0 \end{array}\right),
\end{equation}
so
\begin{equation}\label{Pi 0}
\Pi_0=\Psi(\Psi^\dag\Psi)^{-1}\Psi^\dag=\left(\begin{array}{cc}
|0,0\rangle\langle 0,0|+|1,0\rangle\langle 1,0| & 0 \\
0 & 0 \end{array}\right).
\end{equation}
In fact, a simple comparison between (56) in \cite{TianZhu} and
the above (\ref{Z U2}) makes it clear that here $Z$ also
annihilates two states: $|0,0\rangle$ and $|1,0\rangle$, so
(\ref{Pi 0}) can be obtained directly.

By (\ref{shift}) we introduce a shift operator
\begin{equation}
s=\left(\begin{array}{cc} t & 0 \\
0 & 1 \end{array}\right),
\end{equation}
\begin{equation}
tt^\dag=1,\quad t^\dag t=1-|0,0\rangle\langle
0,0|-|1,0\rangle\langle 1,0|.
\end{equation}
The required $U$  is given as follows:
\begin{equation}
V=\left(\begin{array}{cc} (Z/Z_\rho'~')^{1/2}t^\dag & 0 \\
0 & (Z''~'/Z_\rho''~')^{1/2} \end{array}\right),
\end{equation}
\begin{eqnarray}
U'_{*1}&=&\sqrt{2\zeta+\rho^2}\left(\begin{array}{c}
\sqrt{\zeta+\rho^2}\bar{z}_2\bar{z}_1 \\
-\bar{z}_2(Z_1+Z_2+\theta_2+\rho^2) \\
\sqrt{\zeta+\rho^2}\bar{z}_1\bar{z}_1 \\
-\bar{z}_1(Z_1+Z_2-\theta_1) \end{array}\right)(Z
Z_\rho'~')^{-1/2}t^\dag,\\
U'_{*2}&=&\rho\left(\begin{array}{c} z_1(Z_1''+Z_2') \\
-\sqrt{\zeta+\rho^2}z_1 z_1 \\
-z_2(Z_1''+Z_2'+\zeta+\rho^2) \\
\sqrt{\zeta+\rho^2}z_2 z_1
\end{array}\right)(Z''~'Z_\rho''~')^{-1/2}.\label{Up 2}
\end{eqnarray}

\subsection{The $\theta_2<0$ case}

Unlike the $U(1)$ 2-instanton case in \cite{TianZhu}, the (2,2)
element of the matrix $f$ is well-defined on the vacuum
$|0,0\rangle$:
\begin{equation}
f_{22}|0,0\rangle=(2\theta_1+\theta_2+\rho^2)^{-1}|0,0\rangle.
\end{equation}
Note that this $f_{22}$ tends to that of the $U(1)$ 2-instanton
case when $\rho$ tends to zero, as we have expected.

Now we have
\begin{equation}
e^{\theta_2^{-1}B_2\bar{z}_2-\theta_1^{-1}B_1^\dag
z_1}|0,0\rangle=\left(\begin{array}{cc} |0,0\rangle & 0 \\
-\sqrt{\frac{\zeta+\rho^2}{\theta_1}}|1,0\rangle & |0,0\rangle
\end{array}\right),
\end{equation}
and so
\begin{equation}
\Gamma=\left(\begin{array}{cc} \sqrt{\frac{\theta_1}{2\theta_1+
\theta_2+\rho^2}}|0,0\rangle & 0 \\
-\sqrt{\frac{\zeta+\rho^2}{2\theta_1+ \theta_2+\rho^2}}|1,0\rangle
& |0,0\rangle \\ 0 & 0 \\ 0 & 0
\end{array}\right),
\end{equation}
Then the required $U$ is given as follows:
\begin{equation}
V=\left(\begin{array}{cc} (Z/Z_\rho'~')^{1/2}t & 0 \\
0 & (Z''~'/Z_\rho''~')^{1/2} \end{array}\right),
\end{equation}
\begin{equation}\begin{array}{rl}
U'_{*1}&=\left(\begin{array}{c} \sqrt{\frac{\theta_1}{2\theta_1+
\theta_2+\rho^2}}|0,0\rangle \\
-\sqrt{\frac{\zeta+\rho^2}{2\theta_1+ \theta_2+\rho^2}}|1,0\rangle
\\ 0 \\ 0
\end{array}\right)\langle 0,0|+\left(\begin{array}{c} 0
\\ |0,0\rangle \\
0 \\ 0
\end{array}\right)\langle 1,0|\\
&\quad+\sqrt{2\zeta+\rho^2}\left(\begin{array}{c}
\sqrt{\zeta+\rho^2}\bar{z}_2\bar{z}_1 \\
-\bar{z}_2(Z_1+Z_2+\theta_2+\rho^2) \\
\sqrt{\zeta+\rho^2}\bar{z}_1\bar{z}_1 \\
-\bar{z}_1(Z_1+Z_2-\theta_1) \end{array}\right)(Z
Z_\rho'~')^{-1/2}t
\end{array}\end{equation}
and $U'_{*2}$ is the same as given in (\ref{Up 2}).

\section{Discussion}

To end this paper, we will explain here how our method can be used
to deal with an interesting class of $U(2)$ multi-instantons. This
case is quite similar to the elongated $U(1)$ multi-instanton.

We will take the following ADHM data:
\begin{equation}
I=\sqrt{k\zeta+\rho^2}\left(\begin{array}{cc} e_k & 0
\end{array}\right),\quad J=\rho\left(\begin{array}{c} 0
\\ e_1^\dag \end{array}\right),
\end{equation}
\begin{equation}
B_1=\sum_{i=1}^{k-1}\sqrt{i\zeta+\rho^2}e_i e_{i+1}^\dag,\quad
B_2=0,
\end{equation}
where
\begin{equation}
e_i^\dag=(\stackrel{1}{0},\cdots,\stackrel{i-1}{0},
\stackrel{i}{1},\stackrel{i+1}{0},\cdots,\stackrel{k}{0}).
\end{equation}
One can easily check that the above data do form a $U(2)$
$k$-instanton solution, which includes the $U(2)$ 2-instanton
solution in the previous section as a special case. It is the
$U(2)$ analog of the elongated $U(1)$ $k$-instanton
\cite{elongated}, and we may call it the `elongated $U(2)$
$k$-instanton'.

Because the above data always satisfy the condition $[B_1,B_2]=0$,
it is easy to obtain
\begin{equation}
\Pi_0=\left(\begin{array}{cc} \sum_{n=0}^{k-1}|n,0\rangle\langle
n,0| & 0 \\
0 & 0 \end{array}\right),
\end{equation}
and to calculate $\Gamma$ by (\ref{Gamma}). Then there is no
difficulty to work out the instanton configuration.

\section*{Acknowledgments}

Chuan-Jie Zhu would like to thank Prof. Zhu-Jun Zheng and the
hospitality at the Institute of Mathematics, Henan University.

\appendix

\section{ADHM Construction for Noncommutative 't Hooft Instantons}

In this appendix we apply our method to the construction of
noncommutative 't Hooft instantons \cite{Correa, Lechtenfeld}. We
will only discuss the ASD 't Hooft instantons on ASD $\bf{R}_{\rm
NC}^4$ (i.e., $\zeta=\theta_1+\theta_2=0$ while $\theta_{1,2}\neq
0$).

Take the following ADHM data \cite{Corrigan, Osborn}:
\begin{equation}\label{Hooft}
a=\left(\begin{array}{c} \rho\otimes 1_2 \\
-a^n\otimes\bar\sigma_n
\end{array}\right),\quad
\rho=(\rho_1,\cdots,\rho_k),\quad
a^n=\mathrm{diag}(a_1^n,\cdots,a_k^n),
\end{equation}
where $\rho_i$ and $a_i^n$ are constants parameterizing the scale
and the position of the $i$th instanton respectively. By direct
calculation one finds
\begin{equation}
f=r^{-2}-r^{-2}\rho^\dag\phi^{-1}\rho r^{-2},\qquad r^2\equiv r^n
r^n
\end{equation}
where
\begin{equation}
r^n=x^n-a^n\equiv\mathrm{diag}(r_1^n,\cdots,r_k^n),\quad
\phi=1+\rho_i^2 r_i^{-2}.
\end{equation}
Summation of repeated indices is assumed here for both $n$ and
$i$. It is straightforward to obtain
\begin{equation}
1-KfK^\dag=\phi^{-1}\otimes 1_2,
\end{equation}
\begin{equation}
-\Delta'fK^\dag=-\Delta'r^{-2}\rho^\dag\phi^{-1},
\end{equation}
where a tensor product with $1_2$ is omitted in the second
equation.

From the ADHM data (\ref{Hooft}) we see that the matrices $B_1$
and $B_2$ take very simple diagonal form. So we have also
$[B_1,B_2]=0$ and the extra zero modes from (\ref{Gamma}) are just
$k$ coherent states respectively shifted by the complex components
of $a_1^n,\cdots,a_k^n$. Denote these (orthonormal) coherent
states as $|a_i\rangle$, we will have
\begin{equation}
\Gamma=\left(\begin{array}{c}
\mathrm{diag}(|a_1\rangle,\cdots,|a_k\rangle) \\ 0_k
\end{array}\right).
\end{equation}
Then the required $U$ is again obtained by (\ref{U begin}-\ref{U
end}) where $k'=k$:
\begin{equation}
U=\left(\begin{array}{c}
0 \\ \Gamma
\end{array}\right)Y+\left(\begin{array}{c}
1_2 \\
-\Delta'r^{-2}\rho^\dag
\end{array}\right)\phi^{-1/2}\mathcal{S}.
\end{equation}

\end{document}